\documentclass[11pt,twoside]{article} 
\usepackage{acta-info}
\usepackage{euler}
\usepackage{epstopdf}
\newcommand{\vol}{\textbf{2,} 1 (2010) 28--39}   
\newcommand{\amss}{\amssubj{%
68U20, 68U05
}}     
\newcommand{\CRs}{\CRsubj{%
I.6.3
}}   
\newcommand{\keyw}{\keywords{%
simulation, Coloured Petri Nets, parallel computing, raytracing
}}

\begin{document}

\title{%
Using Coloured Petri Nets for design of parallel raytracing environment
}
\maketitle


\twoauthors{%
\href{http://hornad.fei.tuke.sk/kpi/person/korecko/dcicard.php}{\v{S}tefan Kore\v{c}ko}
}{%
\href{http://dci.fei.tuke.sk/}{Department of Computers and} \href{http://dci.fei.tuke.sk/}{Informatics}, FEEI, Technical University of Ko\v{s}ice,
Letn\'{a} 9, 041 20 Ko\v{s}ice, Slovakia 
}{%
\href{mailto:stefan.korecko@tuke.sk}{stefan.korecko@tuke.sk} 
}{%
\href{http://hornad.fei.tuke.sk/kpi/person/sobota/dcicard.php}{Branislav Sobota} 
}{%
\href{http://dci.fei.tuke.sk/}{Department of Computers and} \href{http://dci.fei.tuke.sk/}{Informatics}, FEEI, Technical University of Ko\v{s}ice,
Letn\'{a} 9, 041 20 Ko\v{s}ice, Slovakia
}{%
\href{mailto:branislav.sobota@tuke.sk}{branislav.sobota@tuke.sk} 
}


\short{%
\v{S}. Kore\v{c}ko, B. Sobota
}{%
Using CPNs for design of parallel raytracing environment
}

\begin{abstract}
This paper deals with the parallel raytracing part of virtual-reality system PROLAND, developed at the home institution of authors. It describes an actual implementation of the raytracing part and introduces a Coloured Petri Nets model of the implementation. The model is used for an evaluation of the implementation by means of simulation-based performance analysis and also forms the basis for future improvements of its parallelization strategy. 
\end{abstract}

\section{Introduction}

During the past several years, high-performance and feature-rich PC graphics interfaces have become available at low cost. This development enables us to build clusters of high-performance graphics PCs at reasonable cost. Then photorealistic rendering methods like raytracing or radiosity can be computed faster and inexpensively. Raytracing is one of computer graphics techniques used to produce accurate images of photorealistic quality from complex three-dimensional scenes described and stored in some computer-readable form \cite{Dietrich07MasModelRend,Georg08FlexHPerfRt,Wald06RtVirtReal}. It is based on a simulation of real-world optical processes. One great disadvantage of such techniques is that they are computationally very expensive and require massive amounts of floating point operations \cite{Heir98loadBal,Not97ParProgRt}. Parallel raytracing takes advantage of parallel computing, cluster computing in particular, to speed up image rendering, since this technique is inherently parallel. The use of clusters \cite{Sob07VisClEnv} for computationally intensive simulations and applications has lead to the development of interface standards such as the MPI and OpenPBS.

This paper provides insight into various means of decomposing the raytracing process (based on the free raytracer Pov-ray \cite{wwwPovrayRt}) and describes a parallel raytracing process management simulation. We decided to use Coloured Petri Nets (CPNs) and CPN tools software for the simulation and a performance analysis based on it. The CPNs and CPN tools were chosen because of good simulation-based performance analysis support, familiar formalism and the fact that the tools are available for free. The other reason was that in the case of some more complicated raytracing process management design we can specify its analytical model using low-level Petri nets and use analytical ``tools'' of Petri nets, such as invariants, and others, including our own results in the field of formal methods, for a verification of the model. After that, the analytical model can be transformed to the CPN model suitable for a performance analysis.

\section{Raytracing and its computation model}
In nature, light sources emit rays of light, which travel through space and interact with objects and environment, by which they are absorbed, reflected, or refracted. These rays are then received by our eyes and form a picture.

Raytracing produces images by simulating these processes, with one significant modification. Emitting rays from light sources and tracking them would be very time-consuming and inefficient, because only a small fraction ends up in the eye/camera, the rest is irrelevant. So instead of this, raytracing works by casting rays from camera through image plane (for each pixel of final image) into the scene and tracking these rays. It computes the intersection of the ray with the first surface it collides with, examines the material properties (casting additional rays for refraction/reflection if necessary) and incoming light from light sources in the scene (by casting additional rays from intersection to each source) and then computes the colour of the pixel in the final image \cite{wwwPovrayRt}.
Raytracing belongs to a set of problems that utilize parallel computing very well, since it is computationally expensive and can be easily decomposed. The two main factors influencing the design and performance of parallel raytracing systems, are the computation model and the load-balancing mechanism \cite{Heir98loadBal}.

There are two principal methods of decomposing a raytracing computation: demand-driven and data-driven (or data-parallel), and there are research activities focused on developing a hybrid model trying to combine the best features of the two models \cite{Not97ParProgRt}.
The final product of raytracer by demand-driven parallel raytracing is an
image of $m\times n$ pixels, and since each pixel is computed
independently, the most obvious way of decomposition is to divide the
image into $p$ parts, where $p$ is a number of processors available and
each processor would compute $m\times n/p$  pixels and ideally, the
computation would be $p$ times faster. This approach is called demand-driven parallel raytracing.
A number of jobs are created each containing different subset of image pixels and these jobs are assigned to processors. Input scene is copied to local memory of each processor. Processors render their parts, return computed pixels, get another job if there is any, and in the end the final \mbox{image is composed from these parts \cite{Dietrich07MasModelRend}.}

Main benefits of this approach are easy decomposition and implementation, simple job distribution and control and the fact that a general raytracing algorithm remains unchanged and scales well. The main disadvantage is that the input scene has to be copied to local memory of each processor, which poses a problem if the scene is very large.

Data-driven parallel raytracing approach, also called data-parallel raytra\-cing, splits the in\-put scene into a number of sections (tiles) and assigns these sections to processors \cite{Dietrich07MasModelRend,Georg08FlexHPerfRt}. Each processor is responsible for all computations associated with objects in this particular section, no matter where the ray comes from. Only rays passing through the processor's section are traced. If a ray spawned at one processor needs data from another processor, it is transferred to that processor. The way the scene is divided into section determines the efficiency of parallel computation. Determining the number of rays that will pass through a section of the scene in order to estimate the sections requiring the most processing is one of the hardest problems to overcome. Using the cost function can be helpful.
Main benefit of this approach is that the input scene doesn't have to be copied entirely to each processor, but it is split into sections, so even very large scenes can be processed relatively easy. Main disadvantage is that this approach doesn't scale very well with growing scene complexity and cluster size, because of task communication overhead and ray transfers \cite{Not97ParProgRt}.

\subsection{Parallelization implementation}
\label{sec:ParallelizationImplementation}

For parallel implementation a cluster-based computing system is used. Cluster-based rendering \cite{Sob07VisClEnv} in general can be described as the use of a set of computers connected via a network for rendering purposes, ranging from distributed non-photorealistic volume rendering over raytracing and radiosity-based rendering to interactive rendering using application programming interfaces like OpenGL or DirectX.

For the raytracing itself, a freeware program Pov-ray is used \cite{dpRusyniak08}, and atop of Pov-ray, a front-end performing parallel decomposition and job control is built. Pov-ray is able to render only a selected portion of the picture, so it's very convenient for naive parallelization. Implementation is limited by Pov-ray's capabilities:

\begin{itemize}\addtolength{\itemsep}{-0.6\baselineskip}
	\item only contiguous rectangular section of image can be rendered in one job,
	\item each job requires parsing the scene and initial computations all again,
	\item each Pov-ray job requires whole scene and
	\item program should be able to handle failures of individual nodes.
\end{itemize}

Because of these facts, the program implements demand-driven computation model. For a load balancing, static or dynamic load balancing by tiling decomposition seems to be the best choice. Implementation uses the Message Passing Interface and SPMD program model.

\begin{figure}
\centering
\scalebox{1.0}[1.0]{  
\includegraphics* [scale=0.9]{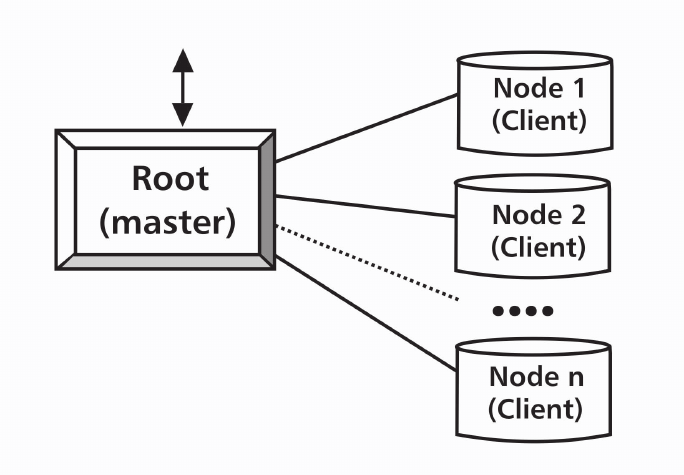}
}
\caption{Basic structure for parallel raytracing implementation}
\label{fig:basicStr}
\end{figure}

Fig. \ref{fig:basicStr} shows the basic used structure. It isn't a typical master/slave scenario, here all nodes are equal, with the exception of the root node, which also controls the whole operation, allocates jobs and interacts with the user. That allows us to utilize massive parallelism.
User puts in the scene to be rendered and additional control information. Root (master) node partitions the final image plane into sections (tiles) and allocates them to nodes. On each node, the process forks and executes Pov-ray to render its part of the image. When finished, it returns the rendered pixels and waits for another job, if required. At the end, the root node puts the whole image together and returns it to the user. It is a simple algorithm. We need to develop a better strategy for distribution of a scene section of rendered image. Better node, time and memory management is necessary. Because the development of an improved strategy using ``real'' hardware and software is expensive and very time-consuming, we decided to use formal CPN models and an appropriate simulation on them instead.

\section{Coloured Petri Nets}
\label{sec:ColouredPetriNets}

Coloured Petri Nets (CPNs) \cite{Jensen07cpnTools} is a discrete-event formal modelling language, able to express properties such as non-determinism and concurrency. It combines a well-known Petri nets formalism with an individuality of tokens to enhance its modelling power and the CPN ML functional programming language to handle data manipulation and decision procedures.

A CPN model has a form of digraph with two types of vertices: places (ellipses) and transitions (rectangles). 

\begin{figure}
\centering
\begin{tabular}{c c c}

\scalebox{0.7}[0.7]{  
\includegraphics* [viewport= 10 10 200 220]{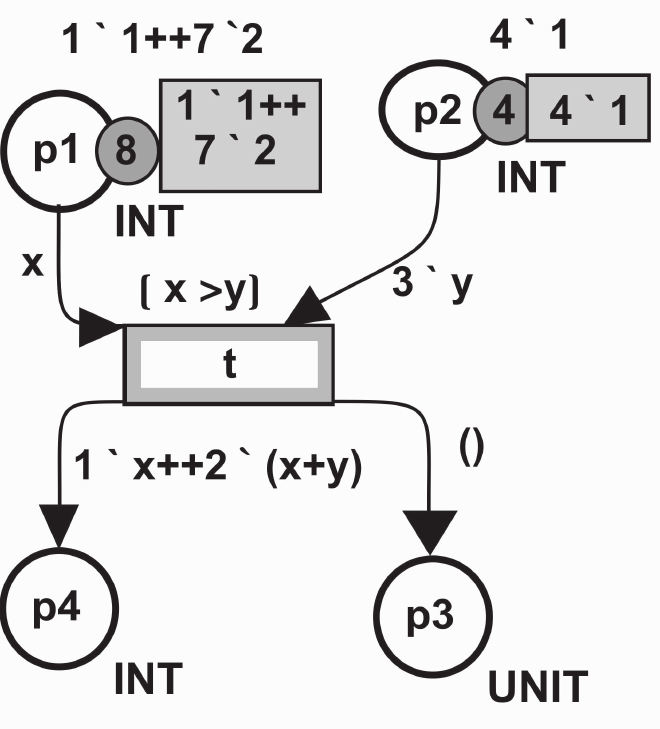} 
}

&
\phantom{aaaaaa}
&

\scalebox{0.7}[0.7]{  
\includegraphics* [viewport= 10 10 200 220]{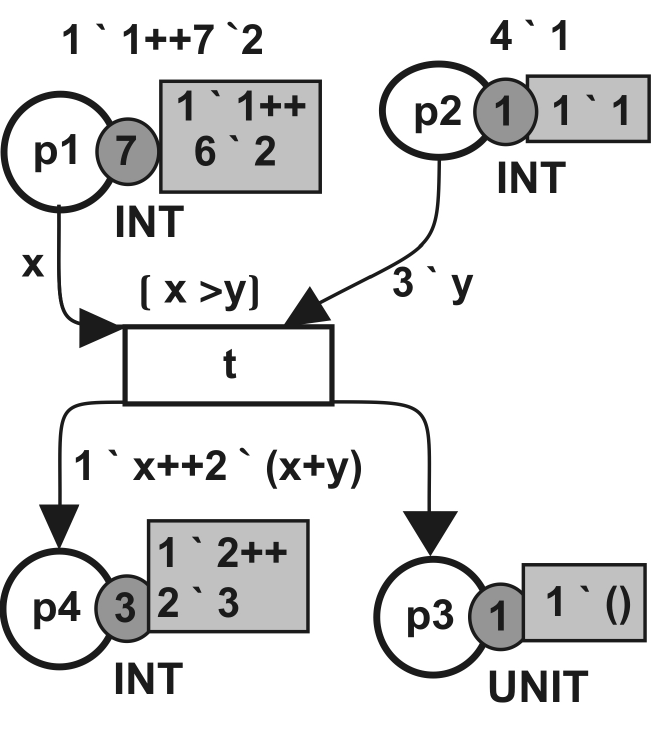} 
}

\\
(a) & & (b)
\end{tabular}

\caption{CPN fragment before (a) and after (b) the firing of the transition $t$}
\label{fig:CPNfragments}
\end{figure}

Each place holds tokens of some type. In CPNs types are called colour sets. Colour sets range from simple ones as UNIT (with the only value ``$()$''), INT, BOOL, to compound sets such as List, Record or Product. An example of user-defined colour sets (record and timed list) can be seen in Fig. 5. Tokens in places define state of CPN, which is called marking. Markings are represented as multisets, i.e. marking ``$1`1++7`2$'' of the place $p1$ from Fig. \ref{fig:CPNfragments} means that $p1$ holds one token of value $1$ and seven tokens of value $2$. If there is only one token in a place, we can omit a number of tokens (for example we can write  ``$4$'' instead of ``$1`4$'').  

Transitions of CPN represent events that change the state (marking) of the net. A transition $t$ can be executed, or fired, when there are enough tokens of corresponding value in places from which there is an arc to $t$. These tokens are removed when $t$ is fired and new tokens are generated in places to which there is an arc from $t$. A number and values of removed and created tokens are determined by corresponding guarding predicates (guards), associated with transitions, and arc expressions. 
A small example in Fig. \ref{fig:CPNfragments} illustrates the behaviour of CPN. The net in Fig. \ref{fig:CPNfragments} has 3 places ($p1$, $p2$, $p4$) of colour set INT and one place of colour set UNIT. Initially the net is in the (initial) marking with ``$1`1++7`2$'' in $p1$ and $4$ tokens of value $1$ in $p2$ (Fig. \ref{fig:CPNfragments}(a)). An actual marking is shown in boxes left to the places. The transition $t$ with the guard ``$x>y$'' can be fired only for $x=2$ and $y=1$ now.  The net after the firing is shown in Fig. \ref{fig:CPNfragments}(b).

\subsection{Performance analysis with CPNs}
\label{sec:PerformanceAnalysisWithCPNs}
To broaden the scope of CPNs usage, facilities allowing simulation-based performance analysis have been added to both the CPNs language and its supporting tool, called CPN tools \cite{wwwCPNtools}. These facilities include time concept for CPNs (timed CPNs), random distribution functions for CPN ML and data collecting and simulation control monitors for CPN tools.

\begin{figure}[!htb]
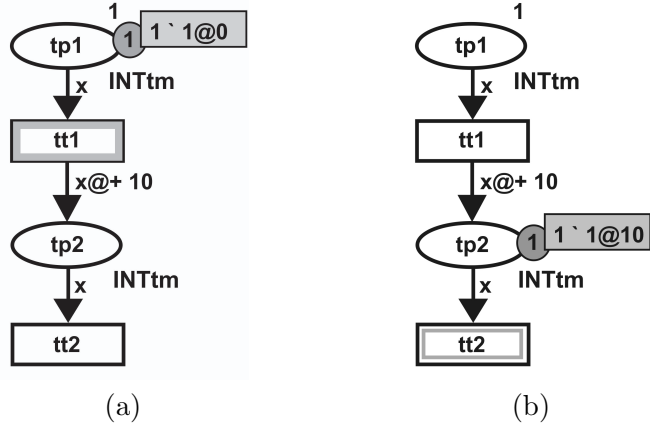

\centering
\begin{tabular}{c c c}
\scalebox{0.7}[0.7]{  
\includegraphics* [viewport= 320 3 455 220]{cpnexbefore} 
}
&
\phantom{aaaaaa}
&
\scalebox{0.7}[0.7]{  
\includegraphics* [viewport= 325 3 453 220]{cpnexafter} 
}
\\
(a) & & (b)
\end{tabular}

\caption{Timed CPN fragment before (a) and after (b) the firing of $tt1$}
\label{fig:tmCPNfragments}
\end{figure}

A (model) time in CPNs and CPN tools is represented as an integer value. There are also values, called time stamps, associated with tokens, representing a minimal time when the tokens are ready for firing. Colour sets of such tokens must be timed. In our models we distinguish timed colour sets by a postfix ``$tm$'' or ``$Tm$''. The model time doesn't change while there is some transition that can be fired. When there is no transition to fire, the time advances to the nearest time value with some transitions to fire. All time-related information in markings, expressions and guards is prefixed by ``$@$''. A small example of timed CPN can be seen in Fig. \ref{fig:tmCPNfragments}(a). Both places can hold timed integers. In the initial marking we have one token of value $1$ and timestamp $0$ in $tp1$. So, $tt1$ can fire in $time = 0$. After firing of $tt1$ one token of value $1$ and time stamp = firing time +10 appears in $tp2$ (Fig. \ref{fig:tmCPNfragments}(b)). In addition, the model time advances to $10$, because there is nothing to be fired in $time=0$ and the token in $tp2$ will not be available (ready) before $time=10$.

Because of space limitations we described CPNs very briefly here. An interested reader can find more information in \cite{Jensen07cpnTools,Wells06cpnPerf} or at \cite{wwwCPNtools}.

\section{CPN model for parallel raytracing}
\label{sec:CPNModelForParallelRaytracing}

\begin{figure}[!htb]
\centering
\scalebox{0.56}[0.56]{  
\includegraphics*{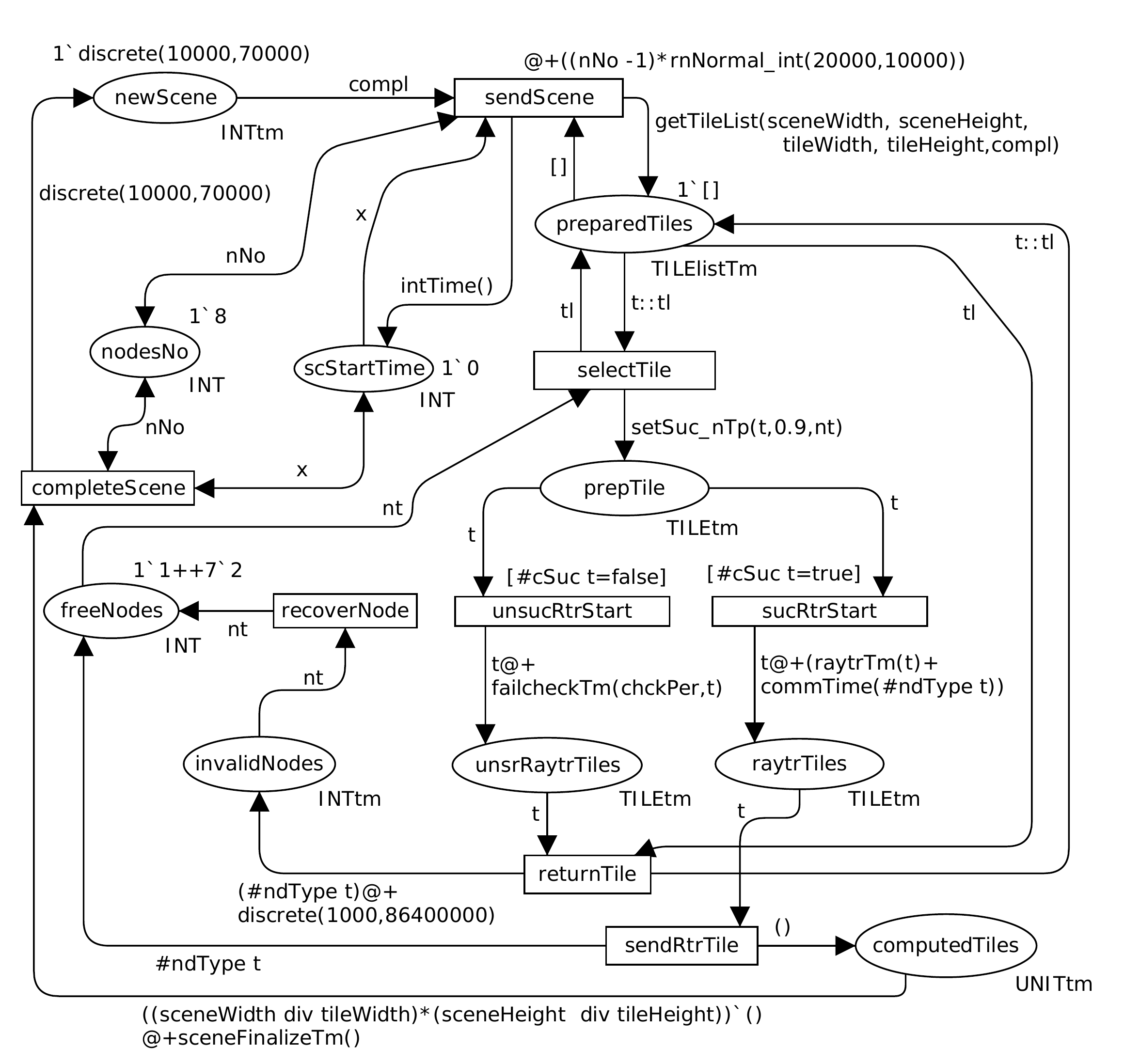}
}
\caption{Timed CPN model of distributed raytracing in 8 computers cluster}
\label{fig:CPNmodel}
\end{figure}

A timed CPN specification of our current implementation of distributed raytracing, as described in section \ref{sec:ParallelizationImplementation}, can be seen in Fig. \ref{fig:CPNmodel}. The time in our model is measured in milliseconds.

In the initial marking there are tokens in places $newScene$, $nodesNo$, $freeNodes$, $scStartTime$ and $preparedTiles$. The place $newScene$ holds one token with randomly chosen value from interval $10000$ to $70000$ (computed by the function \textit{discrete}). This value characterizes a complexity of a scene to be raytraced and its range is based on our practical experience. In general the scene complexity depends on its size, number of objects, objects complexity (number of polygons), objects material (opacity, mirrors, \ldots), illumination model and camera parameters. The place $nodesNo$ holds a token with number of computers in our cluster (8 computers) and $freeNodes$ has one token for each node, where its value designates a type of the node. Albeit all the nodes are equal we have to distinguish between the client nodes (\textit{type 2}) and the master node (\textit{type 1}) that also manages the whole process. So, only about $70\%$ of master performance is used for raytracing. The $preparedTiles$ holds one token with empty list of tiles, because the scene is not divided yet.

Only the transition $sendScene$ can be fired in the initial marking, in $time=0$. Its firing represents sending of the whole scene to each client node. Sending of the scene is a sequential process and its duration is computed by an expression
$$(nNo -1)*rnNormal\_int(20000,10000)$$
where $nNo$ is a number of all nodes and the function \textit{rnNormal\_int(m,v)} returns a value from the exponential random distribution with mean \textit{m} and variance \textit{v}. The firing also divides the scene into the list of tiles with (almost) constant width and height and saves the starting time point of scene raytracing as a token in $scStartTime$. To store information about a tile the colour set TILE is used (Fig. \ref{fig:clsDefs}), where fields \textit{wdt} and \textit{hgt} store tile dimensions, \textit{complxt} stores tile complexity, \textit{cSuc} determines whether tile raytracing will be successful and \textit{ndType} is a type of node where the tile will be raytraced.
\begin{figure}
\centering
\begin{verbatim}
    colset TILE = record wdt:INT *  hgt:INT * complxt:INT 
                     * cSuc: BOOL*  ndType:INT;
    colset TILElistTm = list TILE timed;
\end{verbatim}
\caption{Declarations of some colour sets}
\label{fig:clsDefs}
\end{figure}
The list is generated by the function \textit{getTileList} and is stored as a single token in the place $preparedTiles$. The function also distributes the scene complexity randomly among the tiles. This random distribution is computed by the function \textit{tileCopml} (Fig. \ref{fig:fncDefs}), that is called within \textit{getTileList}. Its first argument, \textit{remTiles}, is a number of remaining tites to be added to the generated list and \textit{remCmpl} is a complexity to be distributed among remaining tiles.
\begin{figure}
\centering
\begin{verbatim}
    fun tileCopml(0, remCmpl) = 0  |
        tileCopml(1, remCmpl) = remCmpl |
        tileCopml(remTiles, 0) = 0 |
        tileCopml(remTiles, remCmpl) =
    let 
      val tCmp = (Real.fromInt remCmpl /
                    Real.fromInt remTiles)
      val cmpl=rnNormalr_int(tCmp*0.8,tCmp *0.7)
    in
      if (remCmpl>cmpl) then cmpl else remCmpl 
    end;
\end{verbatim}
\caption{Definition of \texttt{tileCopml} function}
\label{fig:fncDefs}
\end{figure}

A firing of the transition $selectTile$ means an assignment of raytracing job to a free node \textit{nt}. The selected tile \textit{t} is removed from the list in $preparedTiles$ and moves to $prepTile$.  In addition, the function \textit{setSuc\_nTp} assigns a node type (field \textit{ndType}) to \textit{t} and randomly chooses a raytracing job success (\textit{cSuc}) for \textit{t}. We assume that $90\%$ of all jobs on client nodes will be successful and that the master node never fails. If the field \textit{cSuc} of \textit{t} is true (i.e. ``$\#cSuc\, t=true$'' in CPN ML), then a firing of $sucRtrStart$ moves \textit{t} to $raytrTiles$. The timestamp of $t$ is also increased by raytracing time and a communication delay. The raytracing time is computed by \textit{raytrTm} from all fields of \textit{t} except \textit{cSuc}. The communication delay, computed by \textit{commTime}, is taken from an exponential random distribution and represents the time needed to contact the master node, which can be busy performing other tasks, and to send the raytraced tile to it. After raytracing $sendRtrTile$ moves the tile into already computed ones ($computedTiles$) and frees the node used. 

The path of a fallen one begins with a firing of $unsucRtrStart$, which moves \textit{t} to $unsrRaytrTiles$. The delay computed by \textit{failcheckTm} is a time needed to detect that a given node failed and is not responding. The response of nodes is checked regularly in our implementation, so the delay computed is a randomly chosen multiple of checking period (\textit{chckPer}) with some upper limit. Next, a firing of $returnTile$ moves \textit{t} back to the list in $preparedTiles$ and the failed node to $invalidNodes$, where it waits for recovery. We optimistically suppose that each node recovers within one day. Finally the node is returned to $freeNodes$ by a firing of $recoverNode$.

After successful processing of all tiles the scene can be finalized and the transition $completeScene$ fired. Its firing removes all tokens from $computedTiles$ and generates a new one in $newScene$, so a raytracing process can start over again. There is a data collecting monitor, which saves information about raytracing duration and number of used nodes into the text file for further processing when $completeScene$ is fired.

\section{Simulation experiments}
\label{sec:SimulationExperiments}

To evaluate our implementation of parallel raytracing under various conditions we carried out several simulation experiments on the CPN model created. Here we present results concerning the relation between number of nodes in the cluster and duration of scene raytracing. In these experiments we fixed the scene complexity to 36500 and used a big scene with $30000\times 22500$ pixels and a small one with $10000\times 7500$ pixels. Tile dimensions were $1000\times 750$ pixels in both cases. We considered two scenarios:
\begin{itemize}\addtolength{\itemsep}{-0.6\baselineskip}
\item 
an ideal scenario, where all nodes are equal (i.e. the master can use all of its performance for raytracing) and no computation fails and
\item 
a real scenario, with conditions as described in Section \ref{sec:CPNModelForParallelRaytracing}.
\end{itemize}

\begin{figure}
\centering
\scalebox{0.85}[0.85]{  
\includegraphics*[scale=0.76, viewport=66 0 700 260]{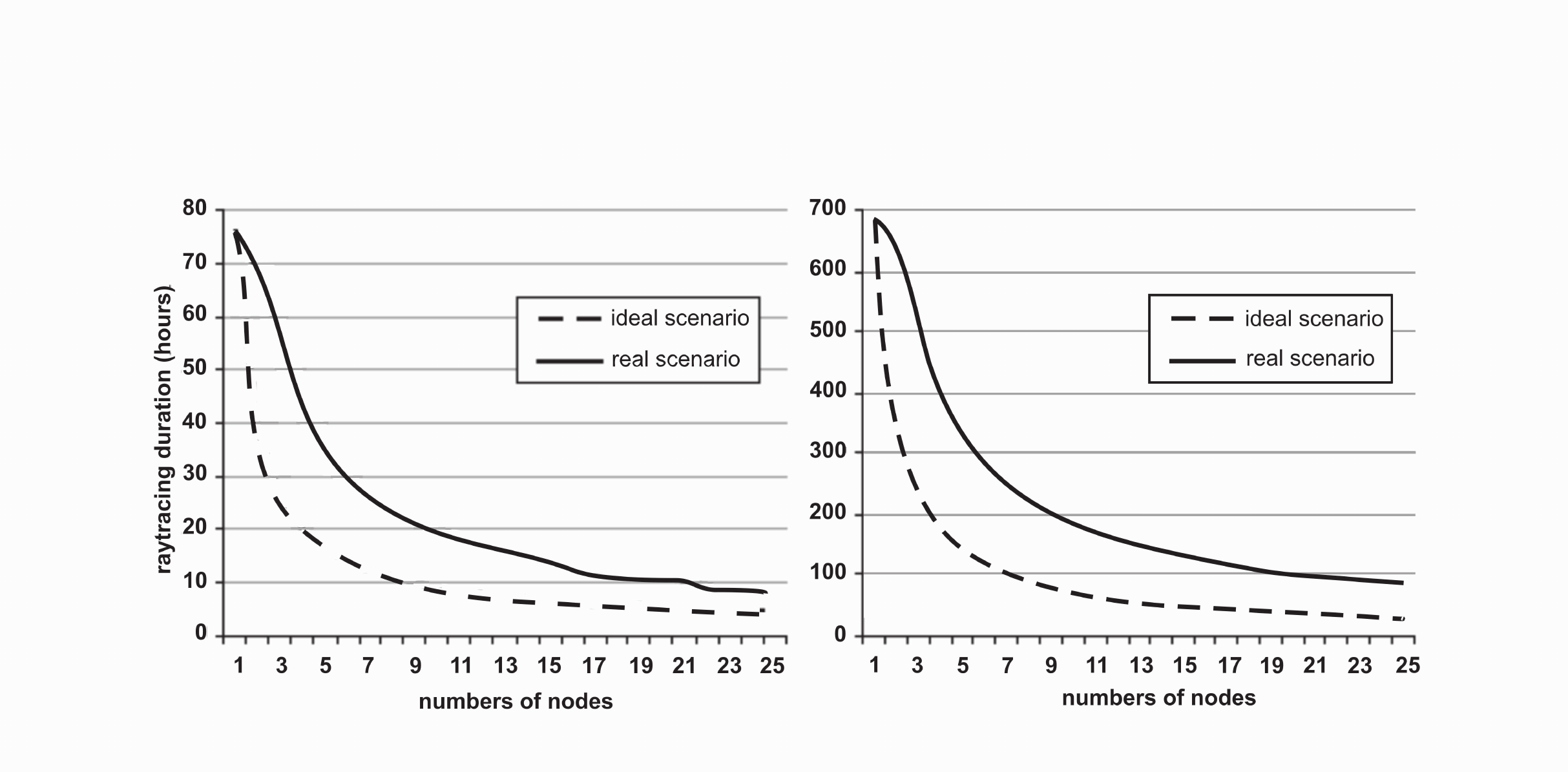}
}
\caption{Number of nodes to raytracing duration ratio for $10000\times 7500$ pixels scene (left) and $30000\times 22500$ scene (right)}
\label{fig:grafy}
\end{figure}

Number of nodes ranged from 2 to 25.
The results obtained are depicted in Fig. \ref{fig:grafy}. As a reference we also included the raytracing duration when only one node is used. The values used in graphs are averages from multiple simulation runs. Of course, in the real scenario, the raytracing time is longer  and the curve is not so ``smooth'' as in the ideal scenario. This is because in the real scenario some nodes can be invalid and raytracing time can be equal or even longer as in the cluster with fewer nodes. The results also reveal that it is not effective to use more than ten nodes in our current implementation of parallel raytracing environment.

\section{Conclusion}
\label{sec:Conclusion}

In this paper we presented our current implementation of distributed raytracing in a cluster environment. We also introduced a CPN model of the implementation, which has been used to evaluate a performance of the implementation and will be used as a basis for the development of an improved parallelization strategy. Our intention is to evaluate possible improvements on corresponding CPN models and choose the best with respect to the performance analysis. In the case of some more complicated strategy we can specify an analytical model using low-level Petri nets first and use analytical facilities of Petri nets, such as invariants and others, including our own theoretical and practical results \cite{SKor08pnTool}, for a verification of the model.

\section{Acknowledgements}
\label{sec:Acknowledgement}

The work presented has been supported by VEGA grant project No.1/0646/09: ``Tasks
solution for large graphical data processing in the environment of
parallel, distributed and network computer systems'' and by Agency of the
Ministry of Education of the Slovak Republic for the Structural Funds of
the EU under the project ``Centre of Information and Communication
Technologies for Knowledge Systems'' (project number: 26220120020).



\rightline{\emph{Received: August 30, 2009  {\tiny \raisebox{2pt}{$\bullet$\!}} Revised February 28, 2010 }}    

\end{document}